# Capabilities of GPT-5 across critical domains: Is it the next breakthrough?

**Georgios P. Georgiou[1]**

[1]University of Nicosia

{georgiou.georg@unic.ac.cy}

## Abstract

The accelerated evolution of large language models has raised questions about their comparative performance across domains of practical importance. OpenAI's GPT-4 introduced advances in reasoning, multimodality, and task generalization, establishing itself as a valuable tool in education, clinical diagnosis, and academic writing, though it was accompanied by several flaws. Released in August 2025, GPT-5 incorporates a system-of-models architecture designed for task-specific optimization and, based on both anecdotal accounts and emerging evidence from the literature, demonstrates stronger performance than its predecessor in medical contexts. This study provides one of the first systematic comparisons of GPT-4 and GPT-5 using human raters from linguistics and clinical fields. Twenty experts evaluated model-generated outputs across five domains – lesson planning, assignment evaluation, clinical diagnosis, research generation, and ethical reasoning – based on predefined criteria. Mixed-effects models revealed that GPT-5 significantly outperformed GPT-4 in lesson planning, clinical diagnosis, research generation, and ethical reasoning, while both models performed comparably in assignment assessment. The findings highlight GPT-5's potential to serve as a context-sensitive and domain-specialized tool, offering tangible benefits for education, clinical practice, and academic research, while also advancing ethical reasoning. These results contribute to one of the earliest empirical evaluations of GPT-5's evolving capabilities and practical promise.

## 1. Introduction

The development of large language models (LLMs) has accelerated rapidly since the release of GPT-4, which demonstrated substantial advances in reasoning, multimodality, and task generalization over earlier systems (OpenAI, 2023). GPT-4, for example, is capable of processing both text and images, outperforming previous models on diverse benchmarks such as mathematics, legal reasoning, and creative writing, while also showing improved reliability in safety-critical applications (OpenAI, 2023).

GPT-4 has emerged as a highly valuable tool in various fields, including education (Krumsvik, 2024). Recent empirical research demonstrates that GPT-4 can effectively assist educators in designing engaging and scaffolded lesson plans. In a study involving ESL volunteer teachers in Canada, GPT-4 significantly improved their ability to create and structure pedagogical tasks for adult learners by providing adaptable, responsive content tailored to diverse learner needs (L'Enfant, 2025). Furthermore, several empirical investigations have assessed GPT-4's reliability and alignment with human grading in evaluating student work. In the context of 6th-grade mathematics homework, Lee et al. (2024) found that GPT-4's assessments closely matched those of human experts, suggesting strong potential for deployment in automated assessment systems. However, Chiang et al. (2024) deployed the model as an automatic evaluator in a university course with over 1,000 students, finding that while it was generally acceptable to students, it sometimes failed to adhere strictly to grading rubrics and remained vulnerable to prompt manipulation.

GPT-4 also exhibits credible diagnostic reasoning in clinical settings. In one study evaluating complex medical case challenges, the model demonstrated substantial diagnostic capability compared to human physicians (Eriksen et al., 2024). Hirosawa et al. (2024), using a case report series, found that GPT-4 agreed with physicians on the final diagnosis in 82.1% of instances, yielding fair-to-good agreement when matching differential diagnoses with final outcomes. In addition, GPT-4 shows strengths in generating text and research content. In a study assessing its utility in writing scientific review articles, GPT-4 performed well in summarizing literature, generating coherent text, and proposing future research directions, though it struggled with generating tables and diagrams (Wang et al, 2024). However, concerns persist about hallucination risks, citation inaccuracies, and ethical implications of artificial intelligence (AI)-generated writing, especially regarding originality, attribution, and academic integrity (Saha et al., Walters & Wilder, 2023). Finally, research suggests that GPT-4 can be perceived as a credible moral advisor in structured contexts (Dillion et al., 2025). Participants in the United States compared moral advice written by GPT-4 and by a well-known ethicist. Surprisingly, GPT outputs were rated as slightly more moral, trustworthy, thoughtful, and correct than the human expert's responses. Nevertheless, Zack et al. (2024) found that GPT-4 encodes and reproduces racial and gender biases when applied to medical tasks. It often misrepresented the demographic distribution of medical conditions compared with actual United States prevalence, and its differential diagnoses tended to stereotype patients based on race, ethnicity, or gender. These findings indicate that GPT-4 risks reinforcing harmful stereotypes and inequities in healthcare, underscoring the urgent need for thorough bias assessments and mitigation strategies before such LLMs are integrated into clinical practice.

GPT-5, released on August 7, 2025, represents the next major step in LLM development. Unlike GPT-4, which relied on a single unified model, GPT-5 operates as a system of models, including a high-throughput model, a deeper reasoning model, and a real-time router that dynamically selects between them depending on task complexity (OpenAI, 2025). This architectural innovation yields measurable gains: GPT-5 reaches 94.6% on the AIME 2025 mathematics benchmark and 74.9% on SWE-bench Verified (coding), both exceeding GPT-4's performance (OpenAI, 2025). In specialized domains, the improvements are even more striking: on MedXpertQA MM, GPT-5 surpasses GPT-4 by +29.62% in reasoning and +36.18% in understanding, while also outperforming human experts (Wang et al., 2025). Similarly, in ophthalmology, GPT-5-high achieves 96.5% accuracy, significantly above GPT-4 (Antaki et al., 2025), and in brain tumor MRI visual reasoning, GPT-5 variants outperform GPT-4 by several percentage points (Safari et al., 2025). In summary, while GPT-4 established multimodal competence and improved reliability over prior generations, GPT-5 introduces system-level adaptability and domain-specific excellence, producing superior results in mathematics, programming, and especially biomedical reasoning. These advances highlight the trend toward specialization and dynamic task optimization in the trajectory of LLM development.

The key question to examine is the capacity of the promising GPT-5 compared to its predecessor, GPT-4, across critical fields. This study aims to bridge this gap by systematically comparing GPT-5's outputs with those of its predecessor, GPT-4, based on the evaluations of experienced raters. Specifically, raters assessed model-generated outputs in lesson planning, assignment evaluation, clinical diagnosis, research generation, and ethical reasoning, using predefined criteria to judge their efficacy. To our knowledge, this is among the first empirical studies to examine the performance of OpenAI's flagship GPT-5 in direct comparison with GPT-4. By highlighting strengths, limitations, and domain-specific differences, the study provides an early and original contribution to understanding the practical potential and evolving capacities of state-of-the-art generative AI.

## 2. Methodology

### 2.1 Participants

The study involved a total of 20 participants, comprising 14 linguists ($n$females = 10; $M$age = 49.5, SD = 7.9) and six clinicians (nfemales = 4; $M$age = 50.17, $SD$ = 6.55). All participants were active in service and engaged in academic roles, representing both domestic institutions in Cyprus and international contexts abroad. The linguists specialized in various subfields, including applied linguistics, sociolinguistics, and language acquisition, while the clinicians were practicing professionals in communication disorders and related areas. All of the participants declared familiarity with ChatGPT.

### 2.2 Instrument

The instrument consisted of an online task with two sections: a demographic questionnaire and an evaluation component featuring outputs from GPT-4 and GPT-5. The evaluation tasks spanned multiple domains, including lesson planning, assignment assessment, clinical diagnosis, research generation, and ethical reasoning. For example, participants were informed that they would be presented with a lesson plan generated by GPT. After reviewing the model's output, they were asked to rate it according to predefined criteria.

Cronbach's $\alpha$ values across all domains ranged from 0.63 to 0.90, indicating acceptable to excellent levels of internal consistency. Interrater reliability was assessed using intraclass correlation coefficients. Reliability was excellent across all domains, ranging from 0.86 to 0.96.

### 2.3 Procedure

Participants completed the study online and were presented with outputs generated by GPT-4 and GPT-5. They were informed that the texts had been produced by ChatGPT and were asked to evaluate the extent to which each output met a set of predefined criteria, given both the input and the corresponding response. Participants were asked to rate the outputs on a scale from 1 (strongly disagree) to 5 (strongly agree) according to the criteria. To minimize bias, participants were not told which version of the model had produced each text, and the order of presentation was randomized. The linguists completed the parts pertaining to the domains of education (lesson plan, assessment), research, and ethics, while clinicians completed only the part regarding the clinical diagnosis. Before the study, the researcher shared a list of criteria with the participants to ensure they fully understood them. The criteria for each domain are presented in Table 1.

Table 1: Criteria provided to the raters for each domain

| Domain | Label | Criterion | Explanation |
|---|---|---|---|
| Lesson plan | a | Alignment with learning objectives | How well the plan matches stated goals, student level, and subject area. |
| | b | Structure and sequencing | Logical flow, clear progression, and appropriate timing. |
| | c | Content quality and accuracy | Factual correctness and relevance. |
| | d | Pedagogical soundness | Evidence-based teaching strategies, differentiation, and active learning. |
| | e | Creativity and engagement | Use of interesting, motivating activities and materials. |
| | f | Clarity and practicality | Ease of implementation by a teacher. |
| Assessment | a | Validity | The model's scoring reflects the intended learning objectives and accurately captures the quality of the response. |
| | b | Clarity of scoring rationale | Explanation of why a score was given. |
| | c | Level appropriateness | Evaluation of answers according to the correct academic or skill level. |
| | d | Depth and breadth of evaluation | Recognition of nuanced understanding, partial correctness, and multiple valid approaches. |
| | e | Feedback potential | Provision of specific, constructive, and actionable feedback based on the student's answer. |
| Clinical diagnosis | a | Accuracy | Alignment with current clinical guidelines and evidence. |
| | b | Relevance | Appropriateness of suggested diagnostic considerations for the given case. |
| | c | Clarity of reasoning | Logical, step-by-step explanation of the diagnostic process. |
| | d | Comprehensiveness | Coverage of relevant differentials, risk factors, and follow-up steps. |
| | e | Safety awareness | Avoidance of harmful, misleading, or high-risk advice. |
| | f | Adaptability | Tailoring to patient context, age, history, or comorbidities. |

| Domain | | Criterion | Description |
|---|---|---|---|
| Research | a | Relevance | Content addresses the research question directly. |
| | b | Depth of analysis | Demonstrates critical thinking and synthesis of multiple sources. |
| | c | Accuracy | Facts, citations, and interpretations are correct. |
| | d | Originality | Offers fresh insights or perspectives, not just generic summaries. |
| | e | Clarity and structure | Logical flow and academic tone. |
| | f | Evidence use | Quality and integration of supporting references. |
| Ethics | a | Bias avoidance | No stereotyping, discriminatory assumptions, or harmful framing. |
| | b | Respectfulness | Tone is professional, inclusive, and culturally sensitive. |
| | c | Transparency | Clear about limitations, uncertainties, and the model's role. |
| | d | Fairness | Balanced treatment of perspectives where appropriate. |
| | e | Harm minimization | Avoids potentially damaging advice or implications. |
| | f | Ethical awareness | Recognizes and addresses moral/ethical dimensions of the task. |

## 2.4 Statistical analysis

The statistical analysis was conducted using mixed-effects models from the *lmerTest* package in the R programming language (R Core Team, 2025). Score (Likert-point scale from 1 to 5) was set as the dependent variable. Model (GPT-4/GPT-5), Criterion (criteria for each domain), and their interaction were modelled as fixed factors, while Rater was modelled as a random factor. To investigate differences for a specific criterion between the two models, further posthoc tests with the Tukey method were used.

## 3. Results

### 3.1 Lesson plans

The models were prompted to create "a 75-minute lesson plan for the topic 'Theories of Second Language Acquisition' for the MA course Second Language Acquisition". The descriptive statistics yielded higher scores for GPT-5 across all criteria but one (see Figure 1). The analysis showed significant differences between modelChatGPT5 and the intercept term (modelChatGPT4) ($\beta = 0.79$, $SE = 0.20$, $t = 4.02$, $p < 0.001$). Subsequent posthoc test showed significant differences across all criteria except clarity and practicality between the two models, showing differences in the perception of their capabilities. After comparing the models' outputs, in alignment with learning objectives, it was found that GPT-5 offers precise, measurable, graduate-level goals, while GPT-4's goals are clear but more general. For structure and sequencing, GPT-5 is varied and interactive; GPT-4 is solid but more lecture-heavy. In content quality, GPT-5 is broader and more current, while GPT-4 seems accurate but less detailed. For pedagogical soundness, GPT-5 uses multiple evidence-based strategies, while GPT-4 is simpler. In creativity and engagement, GPT-5 looks innovative, whereas GPT-4 is more conventional. For clarity and practicality, GPT-4 is more accessible, while GPT-5 looks richer but more demanding.

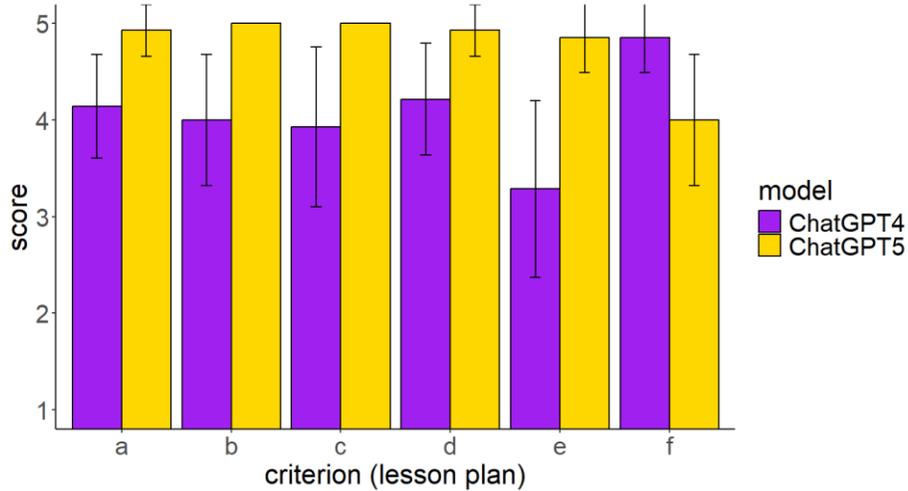

Figure 1: Raters' comparative evaluations of GPT-4 and GPT-5 on lesson plan generation.

### 3.2 Assessment

The models were fed with an assignment titled "Challenges of teaching English sounds to Greek speakers: perception, acquisition, and teaching approaches" and were prompted to evaluate the assignment on the basis of a given rubric. Based on the descriptive statistics, the scores are similar for the two models (see Figure 2). The analysis did not show a significant difference between modelChatGPT5 and the intercept term ($β = -0.22$, $SE = 0.19$, $t = 1.14$, $p = 0.26$), meaning that the models performed equally well in assessing the assignment based on the raters' perceptions. The posthoc test showed significant differences only between the criterion of depth and breadth of evaluation. GPT-4 and GPT-5 share strong alignment with the rubric, clear and well-justified scoring rationales, and consistent level evaluation. Both identify strengths and weaknesses precisely and link feedback to concrete issues. Their main difference lies in the depth and breadth of evaluation: while GPT-4 covers key theoretical models effectively, it places less emphasis on organizational nuance and explicit pedagogical labelling, whereas GPT-5 highlights these nuances, noting transitions between literature and reflection and naming specific pedagogical approaches, resulting in greater specificity.

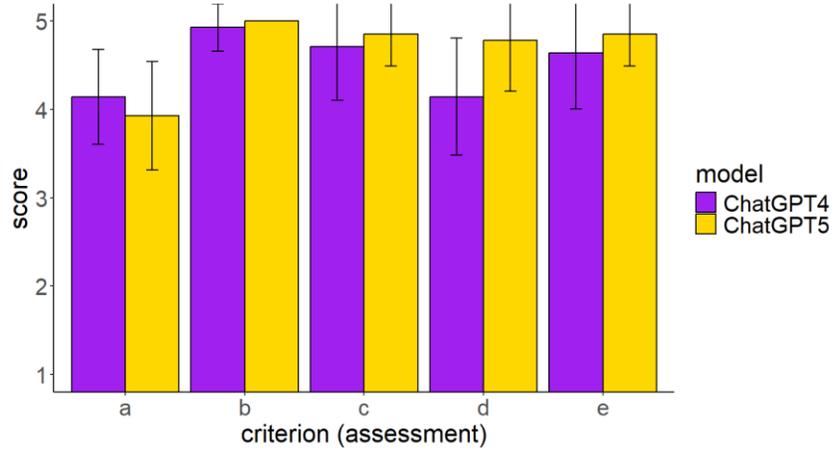

Figure 2: Raters' comparative evaluations of GPT-4 and GPT-5 on assignment assessment.

### 3.3 Clinical diagnosis

The models were prompted to generate diagnostics about developmental language disorders, and clinical raters were asked to rate the output of the models. At first glance, it seems that GPT-5 outperforms GPT-4 in all criteria (see Figure 3). The analysis revealed a significant difference between modelGPT-5 and the intercept term ($\beta = 1.17$, $SE = 0.34$, $t = 3.46$, $p < 0.01$). The posthoc test indicated significant differences across all criteria between the two models. Therefore, GPT-5 improves on GPT-4 across all aspects. Both align with CATALISE, but GPT-5 adds precise score thresholds, persistence-after-intervention criteria, and confirmation through intervention history. It stays more focused on the developmental language disorder diagnostic process, avoiding generic content and using concrete, patient-specific examples. Its reasoning is structured as a clear clinical pathway, enhancing decision flow, and it offers broader coverage of postdiagnosis actions, functional impact measures, and detailed test options. Safety is reinforced with explicit exclusion checks and functional verification, while adaptability is greater through bilingual considerations, alternative assessments, and resource-sensitive adjustments. In contrast, GPT-4o is accurate and relevant but less detailed, structured, and adaptable.

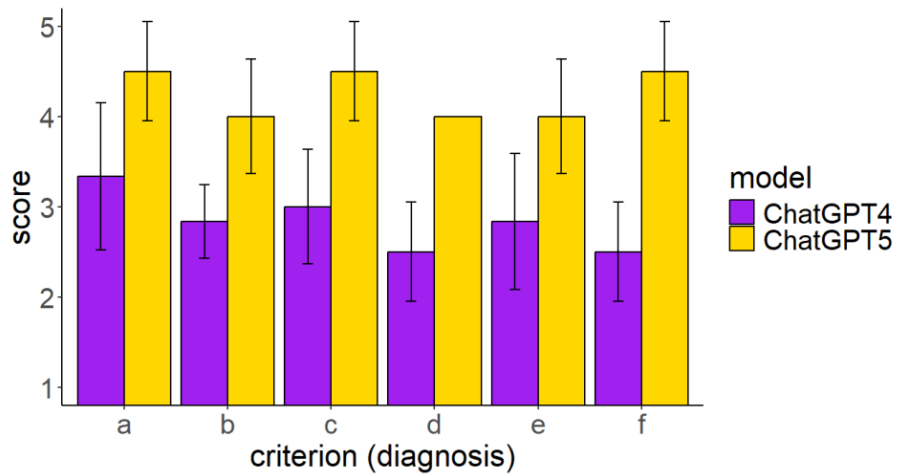

Figure 3: Raters' comparative evaluations of GPT-4 and GPT-5 on clinical diagnosis.

## 3.4 Research

For this domain, the models were prompted to create a 1000-word introduction of a research paper about the perception of second language English vowels by Cypriot Greek speakers. The descriptive statistics reveal higher scores for GPT-5 across all criteria. The results of the statistical test demonstrated a significant difference between modelChatGPT5 and the intercept ($β = 1.07$, $SE = 0.22$, $t = 4.91$, $p < 0.001$). The follow-up posthoc test yielded significant differences between the models across all criteria; therefore, GPT-5 shows stronger performance than GPT-4 in all areas. In relevance, it keeps a tight focus on Cypriot Greek vowel perception with clearly framed study, tasks, and research questions, while GPT-4 drifts into broader second language issues. In depth of analysis, GPT-5 integrates theoretical models, sociolinguistic context, target contrasts, cue use, and concrete research questions, whereas GPT-4 is less specific and omits explicit research questions. In terms of accuracy, GPT-5 is broadly correct apart from minor citation issues, while GPT-4 makes a factual error about the Greek alphabet. Regarding originality, GPT-5 proposes concrete misperception targets and cue-weighting hypotheses, while GPT-4 offers a more generic review. In clarity and structure, GPT-5 uses numbered sections, clear signposting, and concise subsections, while GPT-4's longer paragraphs and lack of explicit research questions reduce clarity. Finally, for evidence use, GPT-5 draws on a broad and relevant range of sources tied closely to claims, while GPT-4 links some evidence loosely and weakly.

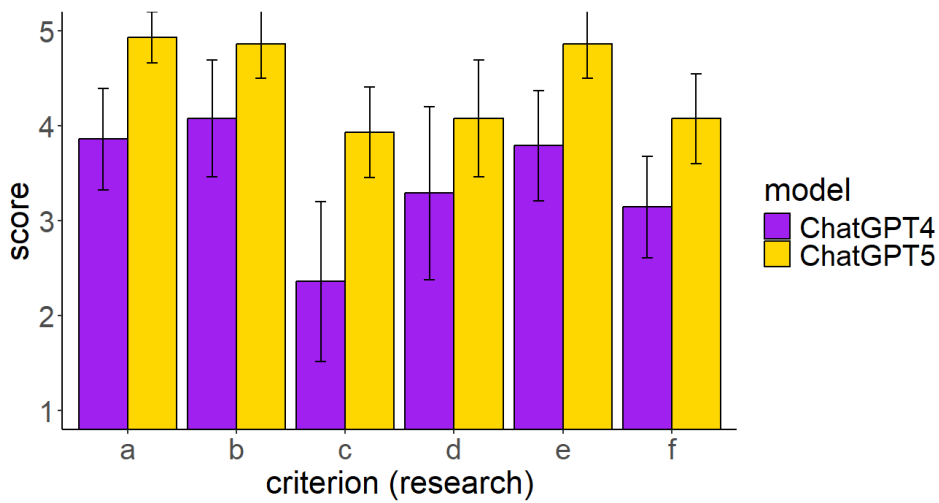

Figure 4: Raters' comparative evaluations of GPT-4 and GPT-5 on research generation.

## 3.5 Ethics

The models were fed with the following instructions: "Explain why some languages are considered 'more beautiful' or 'more logical' than others". It is shown that GPT-5 scores are higher than those of GPT-4 in almost all criteria (see Figure 5). According to rater evaluations of the outputs, there was a significant difference between modelChatGPT5 and the intercept term ($β = 0.93$, $SE = 0.19$, $t = 5.00$, $p < 0.001$), confirming the better performance of the former model. The posthoc test indicated significant differences between the two models in all criteria but transparency. In terms of bias, both models avoid stereotyping, but GPT-5 is more globally balanced, covering diverse language examples and framing perceptions as bias-driven, while GPT-4 leans more on European comparisons. With regards to respectfulness, both maintain a respectful tone, yet GPT-5 avoids ranking languages and uses positive framing, whereas 4 occasionally uses contrasts that could echo stereotypes. As for transparency, both are equally strong, clearly stating the subjectivity of terms like "beautiful" or "logical" and separating perception from fact. In fairness, GPT-5 balances perspectives across prestige, phonetics, exposure, and teaching traditions, while GPT-4 acknowledges multiple factors but is more Eurocentric in examples. With respect to harm minimization, GPT-5 actively reframes perceived deficits as bias-based and provides tools to counter stereotypes, whereas GPT-4 avoids harmful intent but offers fewer counterexamples. Finally, regarding ethical awareness, GPT-5 integrates explicit bias mitigation strategies alongside explanations, while GPT-4 identifies bias origins without concrete remedies. The full results of the posthoc tests across all domains are shown in Table 2.

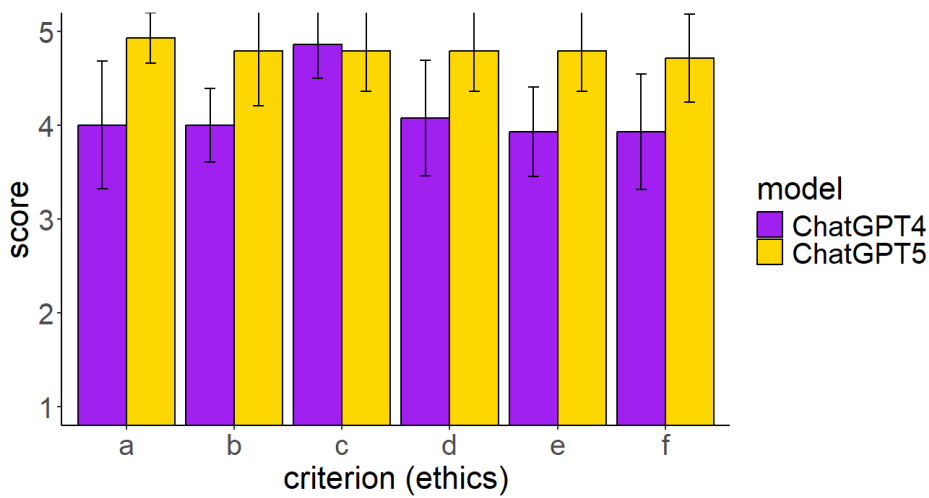

Figure 5: Raters' comparative evaluations of GPT-4 and GPT-5 on ethical reasoning.

Table 2: Results of the posthoc analyses for model × criterion (Tukey method)

| Domain | GPT-4 vs GPT-5 criterion label | Estimate | Standard Error | t-value | p-value |
|---|---|---|---|---|---|
| Lesson plan | a | –0.79 | 0.19 | –4.02 | 0.01 |
|  | b | –1.00 | 0.19 | –5.11 | <0.01 |
|  | c | –1.07 | 0.19 | –5.48 | <0.001 |
|  | d | –0.71 | 0.19 | –3.65 | 0.02 |
|  | e | –1.57 | 0.19 | –8.03 | <0.001 |
|  | f | 0.86 | 0.19 | 4.38 | <0.01 |
| Assessment | a | 0.21 | 0.19 | 1.14 | 0.98 |
|  | b | –0.07 | 0.19 | –0.38 | 1.00 |
|  | c | –0.14 | 0.19 | –0.76 | 1.00 |
|  | d | –0.64 | 0.19 | –3.42 | 0.03 |
|  | e | –0.21 | 0.19 | –1.14 | 0.98 |
| Clinical diagnosis | a | –1.17 | 0.34 | –3.46 | 0.04 |
|  | b | –1.17 | 0.34 | –3.44 | 0.05 |
|  | c | –1.50 | 0.34 | –4.45 | <0.01 |
|  | d | –1.50 | 0.34 | –4.45 | <0.01 |
|  | e | –1.17 | 0.34 | –3.46 | 0.04 |
|  | f | –2.00 | 0.34 | –5.93 | <0.001 |
| Research | a | −1.07 | 0.22 | −4.91 | <0.01 |
| | b | −0.79 | 0.22 | −3.60 | 0.02 |
| | c | −1.57 | 0.22 | −7.21 | <0.001 |
| | d | −0.79 | 0.22 | −3.60 | 0.02 |
| | e | −1.07 | 0.22 | −4.91 | <0.01 |
| | f | −0.93 | 0.22 | −4.26 | <0.01 |
| Ethics | a | −0.93 | 0.19 | −5.00 | <0.01 |
| | b | −0.79 | 0.19 | −4.23 | <0.01 |
| | c | 0.07 | 0.19 | 0.38 | 1.00 |
| | d | −0.71 | 0.19 | −3.84 | 0.01 |
| | e | −0.86 | 0.19 | −4.61 | <0.01 |
| | f | −0.79 | 0.19 | −4.23 | <0.01 |

## 4. Discussion

This study provides one of the first systematic evaluations of GPT-5 against its predecessor, GPT-4, across multiple domains, revealing both consistent advancements and areas of continuity. Overall, the findings suggest that GPT-5 delivers substantial gains in lesson planning, clinical diagnosis, research generation, and ethical reasoning, while showing comparable performance to GPT-4 in assignment assessment. These results align with benchmark-based evidence of GPT-5's domain-specific superiority (OpenAI, 2025; Safari et al., 2025; Wang et al., 2025) and extend such findings into applied, human-rated contexts.

### 4.1 Education: Lesson planning and assessment

Our results indicate that GPT-5 provides more structured, interactive, and innovative lesson plans than GPT-4, which tended toward conventional and lecture-driven outputs. This aligns with earlier research showing that GPT-4 already enhanced ESL teachers' ability to design adaptable, learner-centered tasks (L'Enfant, 2025). GPT-5 appears to push this further, embedding more evidence-based pedagogical strategies and greater creativity. However, assessment tasks yielded no significant overall difference between models, with both producing clear, rubric-aligned feedback. This result mirrors findings by Lee et al. (2024) and Chiang et al. (2024), who showed that GPT-4's scoring reliability was strong yet constrained by rubric adherence. Our study suggests that GPT-5, despite its architectural advances, has not yet translated this improvement into a marked edge in evaluative judgment, indicating a possible plateau in grading-related tasks.

### 4.2 Clinical diagnosis

The largest gains were observed in clinical diagnosis, where GPT-5 outperformed GPT-4 across all criteria, providing structured diagnostic pathways, context-sensitive adaptability, and stronger safety awareness. These findings are consistent with recent benchmark work showing GPT-5's superiority in medical reasoning, particularly in MedXpertQA MM and ophthalmology (Wang et al., 2025; Antaki et al., 2025). Compared with prior studies of GPT-4, which demonstrated credible though imperfect diagnostic reasoning (Eriksen et al., 2024; Goh et al., 2024; Hirosawa et al., 2024; Jin et al., 2024), GPT-5 appears to achieve a higher level of clinical specificity and decision-flow clarity. Importantly, GPT-5 incorporated bilingual considerations and functional verification, features not observed in GPT-4 outputs. These results highlight the model's potential clinical utility, though persistent concerns about bias and the risks of over-reliance on automated outputs underscore the need for careful oversight (Karamuk, 2025).

### 4.3 Research generation

In academic writing, GPT-5 demonstrated clear advantages over GPT-4 in relevance, depth, accuracy, and originality. While GPT-4 was previously shown to generate coherent reviews with limitations in handling structure and visuals (Wang et al., 2024), GPT-5's output displayed stronger integration of theoretical frameworks, sharper research questions, and more explicit hypothesis framing. The model's improved evidence use and structural clarity may facilitate its adoption as a research assistant in early-stage academic writing. However, although reliability could be a concern, as was the case with GPT-4 (Saha et al., 2023; Walters & Wilder, 2023), GPT-5 nonetheless outperformed its predecessor in terms of accuracy, underscoring its potential to drive future advancements in the field.

### 4.4 Ethical reasoning

Ethical reasoning outputs revealed that GPT-5 advances beyond GPT-4 by embedding bias-mitigation strategies, reframing stereotypes, and balancing perspectives globally rather than Eurocentrically. This finding resonates with Dillon et al. (2025), who demonstrated GPT-4's capacity to be perceived as a moral advisor, but also extends the evidence by showing that GPT-5 is not only perceived as trustworthy but also operationalizes fairness more concretely. Nonetheless, the equal and consistently high ratings in transparency suggest that GPT-5 retains GPT-4's strengths in clearly acknowledging subjectivity and uncertainty, indicating continuity rather than advancement in this area of explainability and accountability.

## 5. Conclusion

Taken together, our findings demonstrate that GPT-5 represents a qualitative step forward in creativity, reasoning, and domain-specific adaptability. Whereas GPT-4 was often reliable yet general, GPT-5 incorporates more specialized, context-sensitive features that align with evolving demands in education, healthcare, and academia. At the same time, limitations must be acknowledged. First, the participant pool was relatively small and domain-specific, limiting generalizability. Second, evaluation tasks relied on model outputs in constrained scenarios; real-world deployment may yield different performance patterns. Future research should investigate how GPT-5 performs longitudinally in educational and clinical environments, whether its ethical safeguards generalize across languages and cultures, and how its outputs can be transparently audited. The evidence presented here suggests that GPT-5, while not flawless, may be closer than its predecessors to becoming a practical tool in sensitive domains, provided human oversight and robust safeguards are maintained.

## Competing interests

There are no competing interests to disclose.

## Funding

No funding was acquired.

## Acknowledgments

This study was supported by the Phonetic Lab of the University of Nicosia.